\documentclass[onecolumn]{aastex631}
\usepackage{graphicx}
\usepackage{natbib}
\usepackage{amssymb,amsmath}
\usepackage{savesym}
\savesymbol{tablenum}
\usepackage{siunitx}
\restoresymbol{SIX}{tablenum}
\usepackage{color}

\usepackage{multirow}
\usepackage{mathtools}

\shorttitle{}
\shortauthors{Pallister, Jeffrey \& Stores}

\begin{document}

\title{Exploring the Origin of Solar Energetic Electrons II: Investigating Turbulent Coronal Acceleration}

\author[0000-0002-2376-6725]{Ross Pallister}
\affiliation{Department of Mathematics, Physics \& Electrical Engineering, Northumbria University,  \\
Newcastle upon Tyne, UK\\
NE1 8ST}

\author[0000-0001-6583-1989]{Natasha L. S. Jeffrey}
\affiliation{Department of Mathematics, Physics \& Electrical Engineering, Northumbria University,  \\
Newcastle upon Tyne, UK\\
NE1 8ST}

\author[0000-0002-6060-8048]{Morgan Stores}
\affiliation{University of Minnesota, School of Physics and Astronomy, Minneapolis, 55455, United States of America}
\affiliation{University Corporation of Atmospheric Research, Boulder, Colorado, 80301, United States of America}

\begin{abstract}
Non-thermal particle acceleration in the solar corona is evident from both remote hard X-ray (HXR) sources in the chromosphere and direct in-situ detection in the heliosphere. Correlation of spectral indices between remote and in-situ energy spectra presents the possibility of a common source acceleration region within the corona, however the properties and location of this region are not well constrained. To investigate this we perform a parameter study for both the properties of the ambient plasma of a simulated acceleration region and the turbulent acceleration profile acting on an initially isotropic thermal electron population. We find that the independently varying the turbulent acceleration timescale $\tau_{acc}$, acceleration profile standard deviation $\sigma$ and acceleration region length $L$ result in in-situ spectral index variation of between 0.5 and 2.0 at 1.0 AU for $< 100$ keV electrons. Short timescale turbulent scattering in the flaring corona steepens the spectra by $\sim 0.5$. It was also found that the in-situ spectral index $\delta$ derived from the peak electron flux produces a spectral index $\sim 1.6$ harder than that from a full-flare X-ray photon flux (of spectral index $\gamma$) simulated with the same intermediate parameters. Previous studies have indicated an approximate $\delta \approx \gamma$ relationship for selected flares with measured in-situ electron and X-ray photon observations, suggesting that an extended source region with non-uniform plasma and/or acceleration properties may be necessary to reproduce this relationship.
\end{abstract}

\section{Introduction}\label{sec:introduction}

Solar transient events such as solar flares are highly efficient particle accelerators with observational estimates suggesting a large fraction of the released magnetic energy, at least 10-50\%, is partitioned into energetic particles \citep[e.g., ][]{2012ApJ...759...71E,warmuth2016constraints,2017ApJ...836...17A}. Our understanding of this acceleration mechanism(s) is complicated by the existence of different flare-associated energetic populations, at the Sun and/or released into heliosphere. Whether these populations are produced by the same mechanisms and in the same regions are open questions. Energetic electrons in the range of 1-100 keV are routinely observed in flares by remote sensing X-ray and radio observations, while electrons released into the heliosphere can be observed via their radio emissions and in-situ near Earth and other heliospheric locations with instrumentation onboard Wind \citep{lin_1995}, STEREO \citep{kaiser_2008}, Solar Orbiter (SolO) \citep{muller_2020} and Parker Solar Probe (PSP) \citep{fox2016solar,kasper2016solar}.

A key metric used in analysing energy spectra (flux or fluence) is the spectral index.
Non-thermal acceleration typically results in a non-thermal tail at energies above the base thermal distribution, with shallower gradients or smaller spectral indices (``harder spectra'') typically correlating with a greater degree of non-thermal acceleration. Direct analysis of in-situ electron spectra has been widely performed, including for sub-100 keV energy ranges \citep[e.g. ][]{mewaldt2005proton,rodriguezgarcia_2023,lorfing2023solar}.

Previous studies such as \citet{2007ApJ...663L.109K}, examined flares between 2002-2005, during which, 57\% of Wind events (i.e., the detection of flare-associated energetic electrons) had corresponding hard X-ray (HXR) emission from energetic electrons ``at the Sun'', i.e., those electrons moving on newly formed closed flare loops towards the dense solar chromosphere \citep[e.g. ][]{holman2011implications,kontar2011deducing}. For so-called `prompt’ events, where a time correlation is found between in-situ events and X-ray bursts at the Sun, \cite{2007ApJ...663L.109K} established a correlation between the non-thermal spectral indices of HXR emission observed at the chromosphere and that of electrons detected in-situ in the heliosphere. This indicates a connection between the acceleration mechanisms and/or plasma conditions experienced by coronal electrons during flaring events, and potentially a common source region for both observed populations. To explore this possibility, simulations can be used to produce artificial data for both chromospheric HXR and in-situ electron spectra. By initialising an electron population in a shared ``acceleration'' region, both datasets can be subject to the same acceleration mechanisms and plasma environment. 

In Paper I \citep{pallister_2023} a parameter search for the ambient plasma properties of a coronal acceleration region was conducted. The purpose of this was to first focus on how the plasma environment itself affected electron ejecta into the heliosphere independent of non-thermal acceleration. A beamed non-thermal electron population was injected into a short, hot, dense coronal region and subsequently transported through a cold, sparse heliospheric region out to 1.0 AU. The population was subject to both collisional and non-collisional scattering effects, in addition to adiabatic focusing by a guiding $B$-field modelled as the interplanetary Parker spiral. The electron fluence and peak electron flux distributions were studied at multiple positions through Sun-Earth space for appropriate diagnostics, making note of changes in these diagnostics subject to varying coronal region length $L$, temperature $T$ and electron number density $n$. It was found that for sufficiently large and dense coronal acceleration regions the temperature could be constrained by the maximum electron fluence and peak flux values below 20 keV even with instruments at $\approx 1.0$ AU and energy resolutions as low as 3 keV, comparable to data obtained by recent in-situ instruments on board spacecraft such as SolO.

The purpose of constraining the parameters of a source acceleration region is to better identify the location within the corona where this acceleration takes place. We expand the approach outlined in Paper I to the properties of the acceleration itself. To this end, we change the initial electron population from beamed and non-thermal to isotropic and thermal (based on the ambient plasma temperature). In addition to collisional and non-collisional scattering and adiabatic focusing effects, we also apply turbulent acceleration to better investigate how properties of the surrounding region affect the process of acceleration itself rather than its impact on an imposed pre-accelerated population. If so, we can determine whether certain plasma environments are prohibitive to producing non-thermal distributions typically detected in-situ. Solar flare observations using EUV spectral line observations suggest the presence of extended regions of turbulence, via non-thermal line broadening \citep[e.g. ][]{2021ApJ...923...40S,2017PhRvL.118o5101K}, and in recent years, the dynamic nature of the flare environment has favoured stochastic acceleration mechanisms \citep[e.g.,][]{petrosian2012stochastic} such as magnetohydrodynamic (MHD) plasma turbulence \citep{1995ApJ...438..763G}.

In Section \ref{sec:methods}, we summarise the transport model outlined in greater detail in Paper I \citep{pallister_2023} and expand it to include a turbulent acceleration term. In Section \ref{sec:results}, we study how the energy spectra of coronal ejecta change with varying properties of turbulent acceleration, alongside changing plasma properties for an initially thermal population. In Section \ref{sec:helio} we propagate these ejecta populations out to 1.0 AU and investigate how their spectral indices evolve across the heliosphere. Finally in Section \ref{sec:helio_plus_hxr} we pair a simulated in-situ dataset with a simulated chromospheric HXR distribution that shares a source region with equivalent plasma and acceleration parameters. 

\section{Coronal and heliospheric transport model}\label{sec:methods}

Building upon \citet{pallister_2023}, we model electron acceleration and transport in the corona and heliosphere. The model is composed of two discrete domains with uniform plasma parameters: a hot and over-dense coronal `flaring' region of length $L$ with given electron temperature $T$ and density $n$, and a sparser cold plasma ($T$, $n$ $\approx 0$) representing the wider heliosphere for $z > L$, where $z$ is the position along a magnetic field line extending from $R_\Sun$ to 1.0 AU. We define $z=0$ as the lowest boundary of the acceleration region in the corona. The spatial domains we use are identical to those in Figure 1 of \citet{pallister_2023}.

In each region, electron acceleration and transport are modelled using a Fokker-Planck equation \citep[e.g.,][]{1969lhea.conf..111R,1981phki.book.....L,1986CoPhR...4..183K} by evolving an electron distribution function $f(t,z,v,\mu)$ in time $t$, space $z$, speed $v$ and cosine of the pitch-angle ($\beta$) to the guiding magnetic field $\mu=\cos\beta$,

\begin{gather}\label{eq:fp_full}
\begin{rcases}
\frac{\partial f}{\partial t}+ \mu v\frac{\partial f}{\partial z} =
& \underbrace{+ \frac{\Gamma}{2v^{2}} \left[\frac{\partial}{\partial v}\left(2 v G(u) \frac{\partial f}{\partial v}
+4 u^{2} G(u) f\right)\right]}_\text{collisional energy change}\\
& \underbrace{+ \frac{\Gamma}{2v^{3}} \left[ \frac{\partial}{\partial \mu}\left( (1-\mu^2) \biggl [ {\rm erf}(u) -G(u) \biggr ] \, \frac{\partial f}{\partial \mu} \right) \right]}_\text{collisional scattering}\\
& \underbrace{+ \frac{1}{v^{2}}\frac{\partial}{\partial v}\left[v^{2}D(v,z)\frac{\partial f}{\partial v}\right]}_\text{turbulent acceleration}
\end{rcases} (A)\\
\begin{rcases}
& \underbrace{- \,\frac{v(1-\mu^{2})}{2L_{z}}\frac{\partial f}{\partial \mu}}_\text{adiabatic focusing}
\end{rcases} (B)\\
\begin{rcases}
& \underbrace{+ \frac{\partial}{\partial \mu}\left[D_{\mu\mu}\frac{\partial f}{\partial \mu}\right]}_\text{non-collisional scattering} 
\end{rcases} (C)
\end{gather}

%
where $\Gamma = 4 \pi e^4$ln$\Lambda n / m_e^2$, for electron charge $e$, Coulomb logarithm ln$\Lambda$ and electron mass $m_e$. The error function is given by erf$(u)$ and $G(u) = ({\rm erf}(u) - u {\rm erf}(u))/2u^2$, where $u$ is the dimensionless velocity $u = v/(\sqrt{2} v_{th})$ for $v_{th} = \sqrt{k_BT_e/m_e}$. $L_z$ is the focusing length for a magnetic field $B$. $D(v,z)$ and $D_{\mu\mu}$ are diffusion coefficients (described in detail in Sections \ref{sec:results} and \ref{sec:helio}).  The governing equation is identical to the one used in \citet{pallister_2023}, with the addition of a term on the RHS (right-hand side) for modelling turbulent acceleration.

For the purposes of this study we separate the RHS of Equation \ref{eq:fp_full} into three parts: part (A) contains the collisional and turbulent acceleration components, part (B) the adiabatic focusing component, and part (C) the non-collisional scattering component. Within the results and discussion we will outline which parts of Equation \ref{eq:fp_full} were used for generating the electron spectra and elaborate on the included transport effects. Throughout this study, we take the LHS (left-hand side) of Equation \ref{eq:fp_full} to be unchanged in all versions of our transport model.

To allow the evolution of an electron distribution to be modeled in space, energy, and pitch angle to the guiding magnetic field, Equation \ref{eq:fp_full} can be solved numerically by its conversion into a set of time-independent stochastic differential equations \citep[e.g., ][]{1986ApOpt..25.3145G,2017SSRv..212..151S} for $v$, $\mu$, and $z$, as described in detail in similar modelling \citep[e.g., ][]{2014ApJ...787...86J,2023ApJ...946...53S,pallister_2023}.

\subsection{Turbulent acceleration}
In the coronal flaring region, electrons are accelerated out of a thermal plasma using a turbulent acceleration model first described in \citet{stackhouse2018spatially}, and then further used in \citet{2023ApJ...946...53S}, where electrons are accelerated over an extended acceleration region using a spatially dependent diffusion coefficient,

\begin{equation}\label{eq:turb_disp}
D(v,z)=\frac{v_{\rm th}^{2}}{\tau_{\rm acc}}\left(\frac{v}{v_{\rm th}}\right)^{\chi} \times H(z)
\end{equation}
where $\tau_{acc}$ is the acceleration timescale and $\chi$ is the velocity dependence of the acceleration.

The spatial distribution $H(z)$ is given as a (half-) Gaussian function,

\begin{equation}
H(z)=\exp\left(-\frac{z^{2}}{2\sigma^{2}}\right)\;\;\text{for}\;\; z\ge0.
\end{equation}
with standard deviation $\sigma$. 

In this region, we also account for collisional effects using the `warm-target' model \citep[e.g., ][]{2014ApJ...787...86J,2015ApJ...809...35K}. As in \citet{pallister_2023}, the plasma properties of this 
region are varied following different solar flare observations of (above-the-) loop-top sources \citep[e.g.,][]{2014ApJ...781...43C,2015A&A...584A..89J,2020ApJ...900..192F}, within sensible parameter ranges for the coronal region: temperature $T$ ranging between $10-30$~MK, electron number density $n$ ranging between $10^{9}-10^{10}$ ~cm$^{-3}$ and region size $L$ ranging between $10-40$~Mm.

\subsection{Heliospheric modelling}

In the heliospheric environment, adiabatic focusing and pitch-angle diffusion act in opposite to align or disperse particles (respectively) with regards to the magnetic field, the former being more significant where the magnetic field is strong and the latter where the turbulent mean free path is short. Collisions are taken to be negligible in the sparser heliospheric domain where non-collisional effects have previously been demonstrated to take precedence (due to low heliospheric densities described by empirical models and observations e.g., \citealp{newkirk1967structure,1977SoPh...55..121S,1998SoPh..183..165L,1999SSRv...87..185F,2018SoPh..293..132M}). As with \citet{pallister_2023}, no other acceleration or transport processes are modelled in this study. The heliospheric processes are discussed further in the Section \ref{sec:helio}.

\subsection{Initial, plasma and boundary conditions}\label{sec:init_conditions}

The acceleration region is taken to be a discrete section of the flaring corona with length $L$, uniform plasma temperature $T$ and uniform electron number density $n$, as in \cite{pallister_2023}. The acceleration profile is set as a half-Gaussian peaking at $z=0$ and defined by three additional parameters: the timescale of acceleration, $\tau_{acc}$, the acceleration velocity dependence, $\chi$, and the standard deviation of the half-Gaussian, $\sigma$. 

Each parameter of the acceleration profile is independently varied to investigate their relative impact and effects on the spectra of the electron population ejected into the heliosphere. The tested parameter values are: $\tau_{acc}=[5,10,22,50]$s, $\chi=[2,3,4]$, and $\sigma=[0.1,0.3,0.5,1.0]L$, similar to values used in \cite{2023ApJ...946...53S}. The length, temperature and electron number density of the plasma environment are kept within appropriate ranges for the flaring corona.

Thermal electrons are initialised into the acceleration region with the same half-Gaussian distribution as the acceleration profile. Pitch-angles are initialised as fully isotropic. They are propagated through the acceleration region for 50 seconds with their positions and energies frozen once ejected at $z > L$. This allows for the bulk of the population to eject from the acceleration region into the heliosphere; electrons that cross the $z=0$ boundary are perfectly reflected back into the region.

In the interest of focusing on the low-energy component of the ejecta spectra, results are displayed from $1-100$ keV for the electron fluence (time-integrated flux) and peak of the electron flux (the peak of the flux vs. time plots for a given energy band). This energy range corresponds to that typically used to analyse HXR-producing electrons within the corona and the chromosphere, allowing for direct comparison between the precipitating and escaping electron populations. The electrons themselves are permitted to be accelerated above this limit (to prevent any pile-up at higher energies) but not beyond 300 keV, above which it cannot be assumed that relativistic effects can be neglected.

\section{Coronal acceleration and transport}\label{sec:results}

\subsection{Varying the acceleration parameters}\label{sec:results:cor_acc}

\begin{figure*}[t!]
    \centering
    \includegraphics[width=0.9\textwidth]{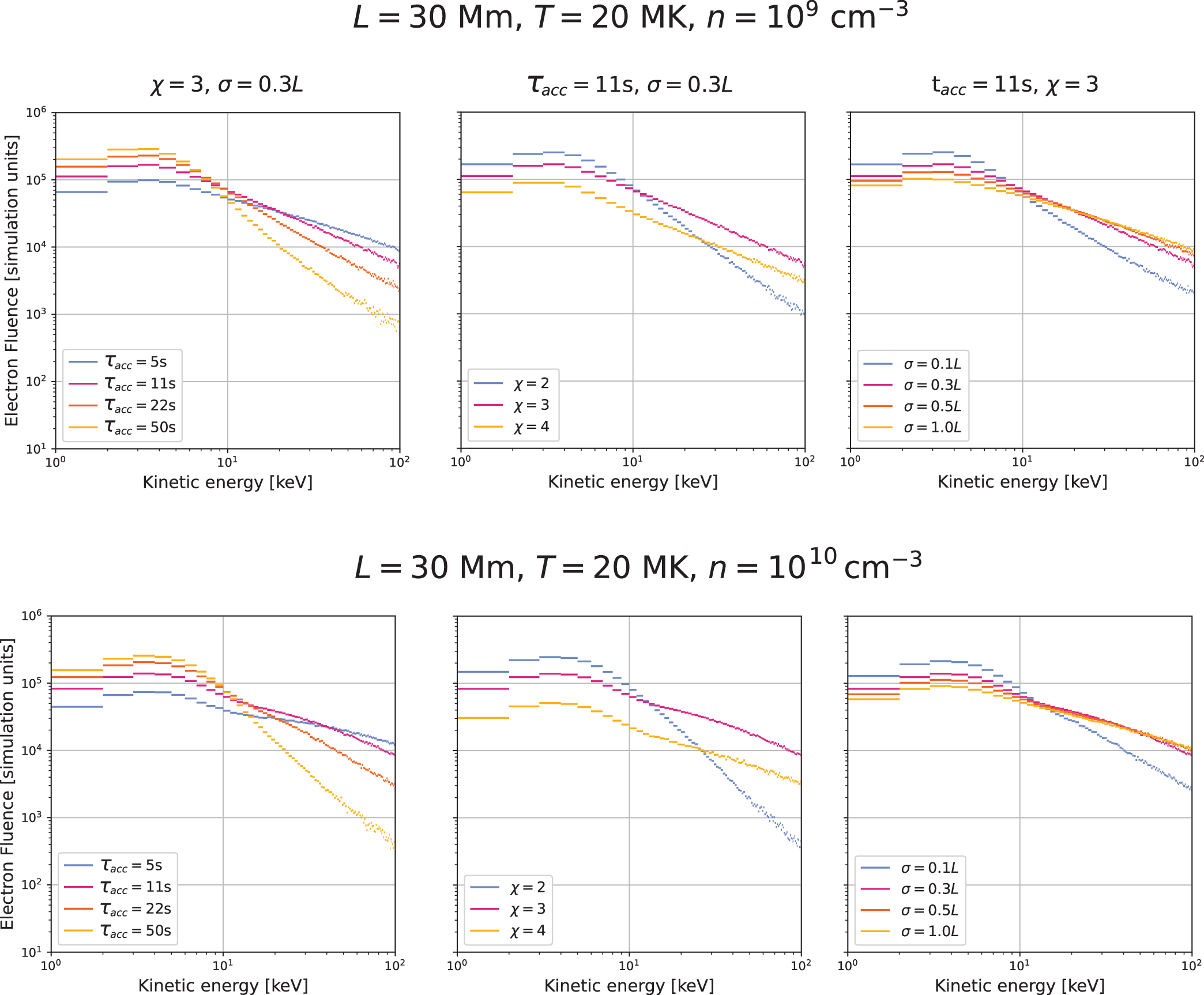}
    \caption{Electron fluence distributions at the boundary of the acceleration region for varying acceleration profile parameters $\tau_{acc}$, $\chi$, and $\sigma$. Where each value is varied, the other two are fixed at intermediate values. The plasma collisional conditions are fixed at $L=30$ Mm and $T=20$ MK with density varied between extremes of $n=10^9$ cm$^{-3}$ (top row) and $n=10^{10}$ cm$^{-3}$ (bottom row). Fluences are binned at 1 keV resolution.}
    \label{fig:acc_fluence_comp}
\end{figure*}

We investigate how the inclusion of various transport effects in the acceleration region influences the resulting electron spectra when the relevant parameters are varied within appropriate ranges for the flaring corona (see Section \ref{sec:init_conditions}). In order to first investigate the effects of implementing a turbulent acceleration component on a collisional only region, we transport the electrons through an acceleration region with part A on the RHS of Equation \ref{eq:fp_full}, subject to varying turbulent acceleration profile parameters in Equation \ref{eq:turb_disp}.

Figure \ref{fig:acc_fluence_comp} shows the fluence distributions at the coronal boundary for low ($n = 10^{9}$~cm$^{-3}$) and high ($n = 10^{10}$~cm$^{-3}$) plasma density values and variable acceleration parameters $\tau_{acc}$, $\chi$ and $\sigma$. Where each of these parameters are independently varied, the other two are set at intermediate values (defined as $\tau_{acc} = 11$s, $\chi = 3$, $\sigma = 0.3L$). By comparing a low and high density acceleration region, the degree by which collisional effects impact the spectra could be quickly identified. The region length and temperature are fixed at $L=30$ Mm and $T=20$ MK respectively. In all cases the $1-5$ keV energy range retains the shape of the thermal distribution. At smaller values of $\tau_{acc}$ and larger $\chi$ we see smaller spectral indices between $10-100$ keV as the degree of acceleration is greater, leading to harder non-thermal tails. Increasing $\sigma$ has a similar effect as the acceleration profile is more extended across the region. While the spectral index between $10-100$ keV remains relatively constant for the lower density acceleration region, we note that at higher densities a ``bump"-like feature is observed at about $30-40$ keV for $\tau_{acc} < 22$~s, $\chi = 3$ and/or $\sigma > 0.3L$; parameters consistent with the greatest degrees of acceleration. The result of this is an apparent break in the fluence distribution gradient, with a harder spectral index below the break energy.

\begin{figure*}[t!]
    \centering
    \includegraphics[width=0.9\textwidth]{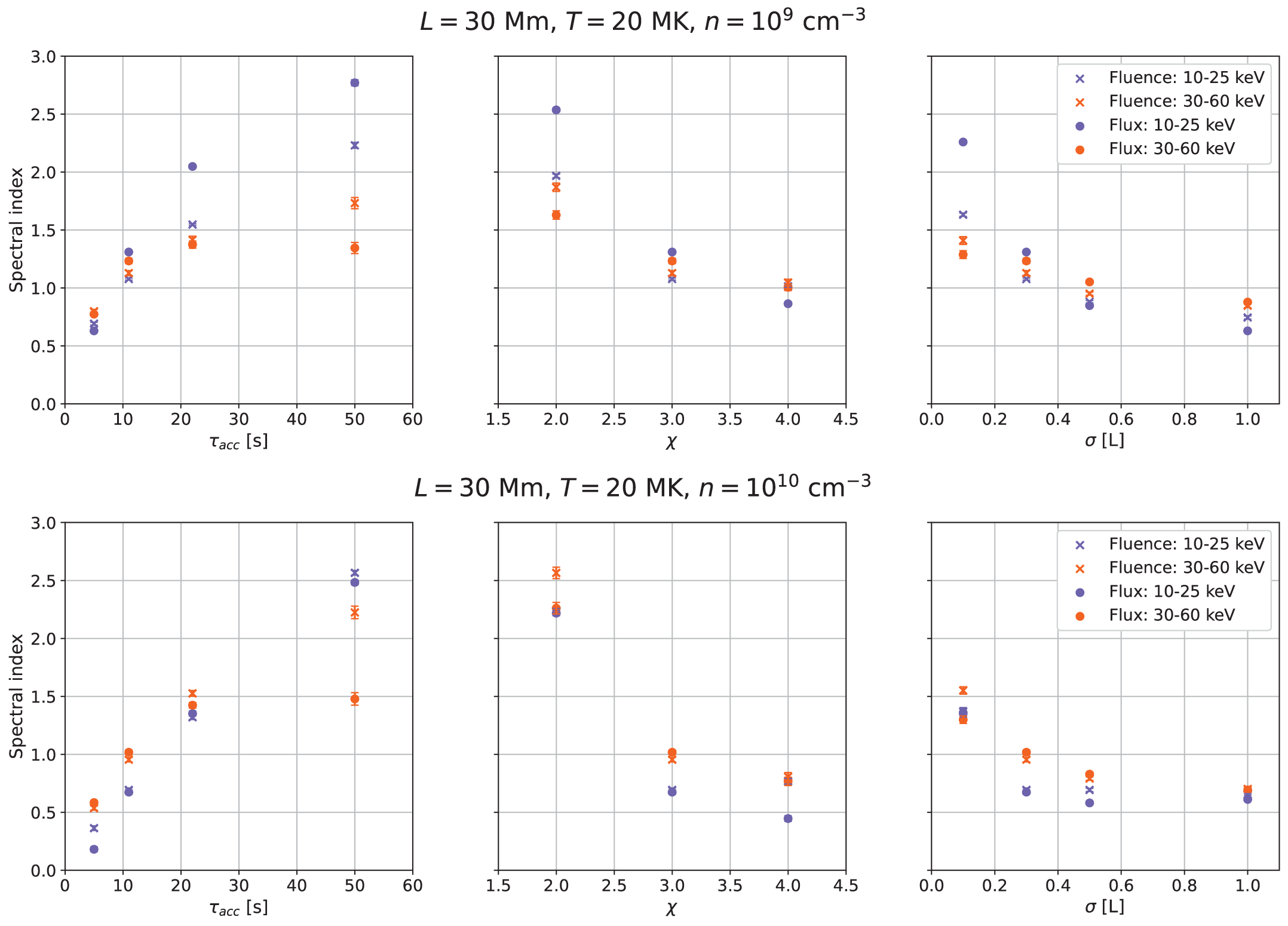}
    \caption{Spectral indices for the fluence (crosses) and peak flux (circles) distributions at the acceleration region boundary for varying acceleration profile parameters $\tau_{acc}$, $\chi$, and $\sigma$. As in Figure \ref{fig:acc_fluence_comp}, other parameters are fixed at intermediate values with the density varying from $n=10^9$ cm$^{-3}$ (top row) to $n=10^{10}$ cm$^{-3}$ (bottom row). The spectral indices are separated into two discrete regions, covering the lower-energy range of [10:25] keV (blue) and a higher-energy range of [30:60] keV (orange).}
    \label{fig:acc_bound_params_comp}
\end{figure*}

The spectral indices for electron fluence and electron peak flux are examined in more detail in Figure \ref{fig:acc_bound_params_comp}. The upper and lower ranges of the two indices are chosen to be $10-25$ and $30-60$ keV, as these are the approximate ranges of the two broken spectral components (if present) identified in Figure \ref{fig:acc_fluence_comp}. In general the spectral index in both fluence and peak flux increases with longer $\tau_{acc}$, lower $\chi$ and shorter $\sigma$, consistent with lower acceleration and distributions closer to thermal. In lower densities both spectral indices are comparable when the population is sufficiently accelerated. Conversely, for higher plasma density we see a broadly higher spectral index at the $30-60$ keV range. The high and low-energy indices become comparable again at very high accelerations as a result of large $\chi$ or $\sigma$, suggesting the spectral break produced by these simulations is an intermediate acceleration effect.

\subsection{Varying the plasma parameters}\label{sec:results:cor_pla}

In addition to studying the individual effects of separable acceleration parameters in a given plasma environment, we also investigate how the inclusion of turbulent acceleration can influence the fluence and flux spectra of different plasma parameter configurations. The approach is similar to that in \cite{pallister_2023}: the physical length $L$ and plasma temperature $T$ are individually varied. As in Section \ref{sec:results:cor_acc}, we initialise an isotropic thermal electron distribution and include a turbulent acceleration term. The tested ranges of these parameters are $L=[10,20,30,40]$ Mm and $T=[10,20,30]$ MK. The acceleration parameters are fixed at intermediate values of $\tau_{acc} = 11$s, $\chi = 3$ and $\sigma = 0.3L$; note that as the value of $\sigma$ scales with $L$, a larger coronal region results in a proportionally more extended Gaussian acceleration profile with the same peak magnitude. As high and low values for the electron number density $n$ have already been investigated in Section \ref{sec:results:cor_acc}, the effect of varying $n$ will not be discussed further here.

\begin{figure*}[t!]
    \centering
    \includegraphics[width=0.75\textwidth]{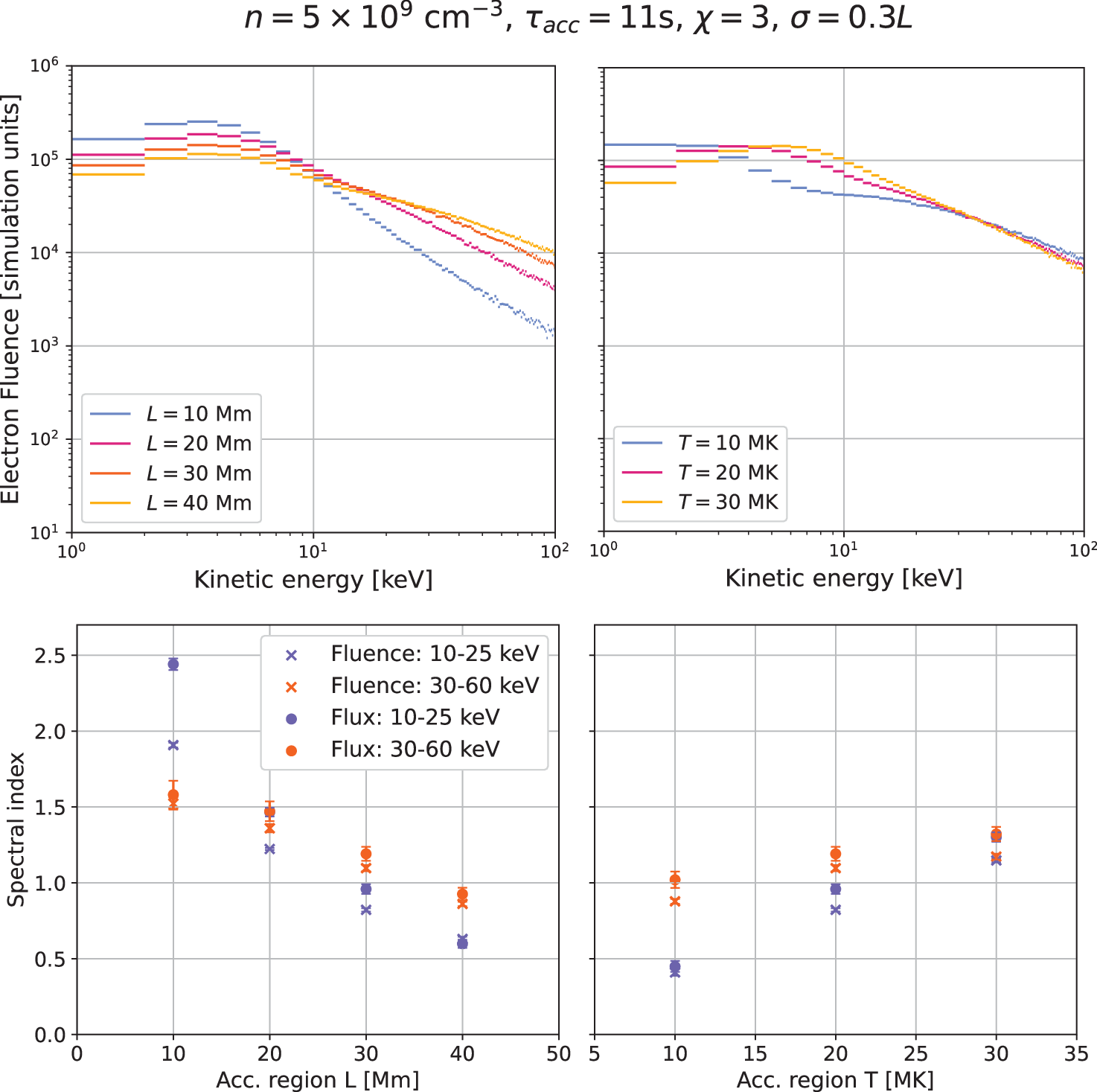}
    \caption{Electron fluence and spectral indices at the acceleration region boundary for varying plasma parameters $L$ and $T$, for fixed intermediate acceleration profile parameters and density $n$. The spectral indices cover the lower-energy range of [10:25] keV (blue) and a higher-energy range of [30:60] keV (orange).}
    \label{fig:pla_bound_params_comp}
\end{figure*}

Figure \ref{fig:pla_bound_params_comp} shows the relation between the plasma parameters and the spectral indices for both fluence and peak flux distributions. As $L$ increases, so does the hardness of the spectra at all energies above 10 keV. Varying $T$ significantly affects the sub-30 keV spectra, as the thermal component of the fluence and peak flux distributions will peak at different energies. At higher $T$ the thermal distribution overlaps the lower range of the non-thermal tail leading to a softening of the spectra (i.e. a steeper gradient). Conversely, at low $T$ the non-thermal tail appears as a clearly separate part of the fluence distribution with its own peak and significant hardening between 10-25 keV.

\subsection{Investigating the non-collisional effects in the acceleration region}\label{sec:results:focusing}

The work in Sections \ref{sec:results:cor_acc} and \ref{sec:results:cor_pla} has been centered around investigating the effects of the interplay between collisional plasma conditions and turbulent acceleration. What has been neglected however is the influence of non-collisional effects: while these are expected to play a more dominant role in the heliosphere where the environment is much sparser and turbulent acceleration does not take place (at least in our model) it is not immediately clear how much it will impact the ejecta distributions.

The first non-collisional parameter we examine is the magnetic field in the acceleration region, which contributes to adiabatic focusing. The adiabatic focusing term in part B of Equation \ref{eq:fp_full} is dependent on the magnetic field strength $B$ to gradient ratio $dB/dz$ \citep[e.g.][]{2009ApJ...693...69D},

\begin{equation}\label{eq:Lz}
    L_z(z) = \frac{B(z)}{\left(-dB/dz\right)}
\end{equation}

Typical magnetic field strengths for flare loops in the solar corona range from the order of $\sim$$100-1000$G close to active region footpoints \citep{yang_2014} to $\sim$$10$G up to $30$Mm above the chromosphere, near the top of flare loops \citep{zhu_2021,brooks_2021}, with some recent estimates up to the range of $300-350$G \citep{kollhoff_2021}. Smaller-scale structures such as coronal jets typically have lower field intensities, in the range of $5-25$G \citep{sako_2014,raouafi_2016}, demonstrating a wide range of $B$-field configurations in the lower corona. How the presence of the aforementioned magnetic field in acceleration regions 10s of Mm above the chromosphere affects coronal electron transport will be investigated using both parts A and B of Equation \ref{eq:fp_full}. Later we will consider the effect of (short timescale) turbulent non-collisional scattering which acts in opposition to the focusing (part C in Equation \ref{eq:fp_full}).

\begin{figure*}[t!]
    \centering
    \includegraphics[width=0.8\textwidth]{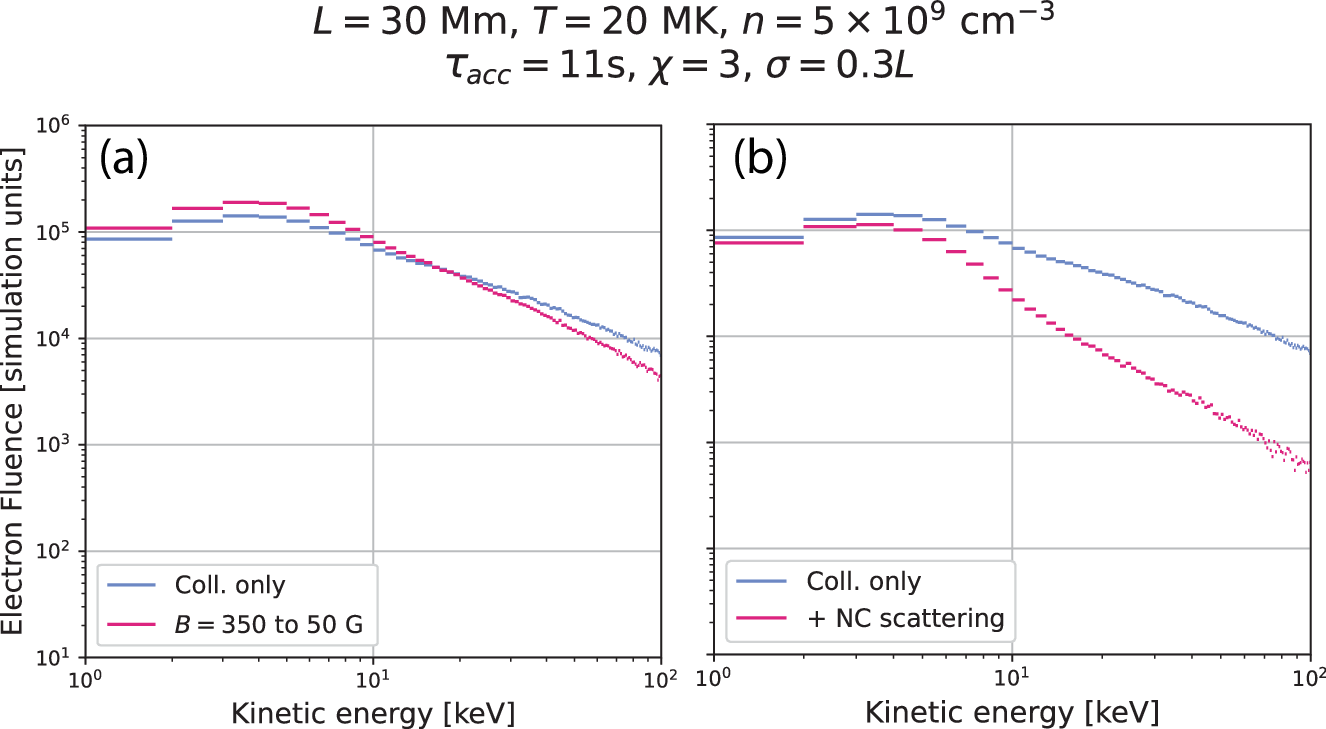}
    \caption{Electron fluence at the acceleration region boundary for varying non-collisional parameters, where plasma-collisional and acceleration profile parameters are fixed at intermediate values. In each plot the blue lines represent a collisional and acceleration only case. (Left) comparison of the fluence spectra for a collisional only region to that with one with adiabatic focused driven by a linear $B$-field, decreasing from 350 G at the bottom of the region to 50 G at the ejection boundary. (Right) fluence variation between a collisional-only region and one with an included non-collisional (NC) scattering term with turbulent scattering mean free path $\lambda_s$ inversely proportional to electron energy E, as in Equation \ref{eq:lambda_acc}.
    }
    \label{fig:NC_fluence_comp}
\end{figure*}

The above observational studies of flaring coronal structures indicate a broad range of magnetic field strength $B$ of order $100-1000$ G close to the chromosphere decreasing to order $1-10$ G at the top of flare loops and coronal jets. The acceleration regions we model are assumed to begin partway up such structures (where magnetic reconnection is expected to take place), so we investigate how the inclusion of adiabatic focusing with an imposed linear $B$-field decreasing as 

\begin{equation}
    B(z) = (B_{\text{min}}-B_{\text{max}}) \cdot \frac{(z - R_\Sun)}{L} + B_{\text{max}} 
\end{equation}
ranging from $B_{\text{max}} = 350$ to $B_{\text{min}} = 50$ G in Equation \ref{eq:Lz} affects the resultant spectra. Note that this field geometry results in a constant magnetic field gradient $dB/dz$ in Equation \ref{eq:Lz}.

The electron fluence distributions for an adiabatically focused spectrum are compared with a collisional-only accelerated spectrum in Figure \ref{fig:NC_fluence_comp}a. The inclusion of magnetic focusing leads to a softening of the spectral index between $10-100$ keV, as electrons are beamed out of the acceleration region more efficiently and are not subjected to turbulent acceleration effects for as long.

We now consider the non-collisional part (C) of Equation \ref{eq:fp_full}, calculated with an isotropic turbulent scattering approximation using the following diffusion coefficient and its $\mu$ derivative \citep[e.g][]{1989ApJ...336..243S,2020A&A...642A..79J}:

\begin{equation}
    D_{\mu\mu} = \frac{v}{2\lambda_s}(1 - \mu^2)
\end{equation}
\begin{equation}
\frac{\partial D_{\mu\mu}}{\partial \mu} = -\frac{ v \mu}{\lambda_s}
\end{equation}
where $\lambda_s$ is the mean free path of the turbulent scattering. In \cite{2018A&A...610A...6M} the value of $\lambda_s$ in the flaring corona was found to follow the inverse electron energy $E$ [keV] relation,

\begin{equation}\label{eq:lambda_acc}
    \lambda_s = \lambda_{s,0} \left(\frac{25 \text{keV}}{E} \right)
\end{equation}
for observed value $\lambda_{s,0} = 2 \times 10^8$ cm.

Figure \ref{fig:NC_fluence_comp}b compares the electron fluences of a collisional and acceleration only case with spectra where non-collisional scattering is included, using parts (A) and (C) of Equation \ref{eq:fp_full}. The inverse-$E$ relationship of $\lambda_s$ shows a softening of the non-thermal energy spectra compared to a collisional-only case, similar to the trend seen in \cite{2023ApJ...946...53S}. We also note a depletion of electron ejecta due to efficient scattering reducing the likelihood of electrons being trapped close to $\mu = 0$, which is more prevalent in a collisional-only scattering case (i.e., the case where the turbulent scattering in the model is assumed to operate on timescales close to the collisional scattering rate).

\section{Heliospheric electron energy spectra}\label{sec:helio}

To examine whether variation in any of the tested parameters could produce differing spectral features detectable by in-situ instruments, we first need to establish that they are preserved out to locations in the heliosphere where data is readily available. It was determined in \citet{pallister_2023} that a non-thermal electron beam injected into a sufficiently dense collisional-only region would retain spectral signatures of coronal thermalisation out to 1.0 AU. Where the spectral index varied little in the previous study due to a uniform pre-accelerated initialisation between parameter sets, it is now expected to vary significantly due to the inclusion of the turbulent acceleration component with its own parameters, as seen in the ejecta spectra in Section \ref{sec:results}. To this end, we perform another parameter search for spectra between the upper boundary of our acceleration region and 1.0 AU. 

We model a heliospheric environment as a region extending from the upper boundary of the acceleration region to 1.0 AU, focusing on interplanetary space between the Sun and Earth. After ejecting from the acceleration region the population enters a region typically referred to as the middle corona between 1.5 and 6 $R_\Sun$ \citep{west_2023}. Observations and modelling of this region typically find $n$ to be 1 to 3 orders of magnitude smaller than our minimum tested value. As such, we do not expect heliospheric collisional effects to significantly affect the ejecta spectra. This leaves us with only parts (B) and (C) of Equation \ref{eq:fp_full} to model the heliospheric transport.

In the heliosphere, $D_{\mu\mu}$ and its $\mu$ derivative are modelled as \citep[e.g.][]{2009ApJ...693...69D,2013JSWSC...3A..10A}:

\begin{equation}
    D_{\mu\mu} = K (1 - \mu^2) (|\mu|^{q-1} + h)
\end{equation}
\begin{equation}
\frac{\partial D_{\mu\mu}}{\partial \mu}= K\mu\bigg[(q-1)(1-\mu^{2})|\mu|^{q-3} - 2(|\mu|^{q-1} + h)\bigg]
\end{equation}
where

\begin{equation}
    K = \frac{3 v}{2 \lambda_s (4-q) (2-q)}
\end{equation}
where $q=5/3$ is the spectral index of magnetic field fluctuations, taken as a Kolmogorov spectrum, and constant $h=0.01$, which is a corrective factor representing nonlinear scattering effects that are dominated at $|\mu| > 0$ in the heliosphere.

The value of $L_z$ in Equation \ref{eq:Lz} across the heliosphere is derived from the magnetic field given by Equation A2 in \citet{2023ApJ...956..112K},

\begin{equation}\label{eq:B_field_kontar}
    B(z) = 0.5 \frac{\left(\frac{z}{R_\Sun} - 1\right)^{-\frac{3}{2}}}{\left(\frac{z}{10 R_\Sun} + 1\right)} + 1.18 \left(\frac{R_\Sun}{z}\right)^2
\end{equation}

Equation \ref{eq:B_field_kontar} combines the models of \citet{1978SoPh...57..279D}, which well describes the magnetic field below $10 R_\Sun$, and \citet{1987SoPh..109...91P}, which better matches interplanetary space from $10 R_\Sun$ to 1.0 AU. 

Following \citet{2018PhDT.......143A}, the mean free path $\lambda_s$ in $D_{\mu\mu}$ is now calculated as in Equation \ref{eq:lambda_helio},

\begin{equation}\label{eq:lambda_helio}
    \lambda_s = \lambda_{s,\Earth} \Bigg(\frac{z}{z_\Earth}\Bigg)^\kappa \Bigg(\frac{p}{p_{\text{min}}}\Bigg)^{2 \xi}   
\end{equation}
dependent on the mean free path at 1.0 AU, $\lambda_{s,\Earth} = 0.3$ AU, the ratio of electron position $z$ relative to 1.0 AU ($= z_\Earth$) and the ratio of current and minimum electron momenta $p$ (derived from the minimum allowed kinetic energy in the heliosphere). $\kappa$ and $\xi$ are parameters that quantify the degree to which the electron momentum and radial distance from the Sun affect the mean free path. As in \citet{pallister_2023}, we take $\kappa = 0.5$ and $\xi = -0.2$.

\begin{figure*}[t!]
    \centering
    \includegraphics[width=1.0\textwidth]{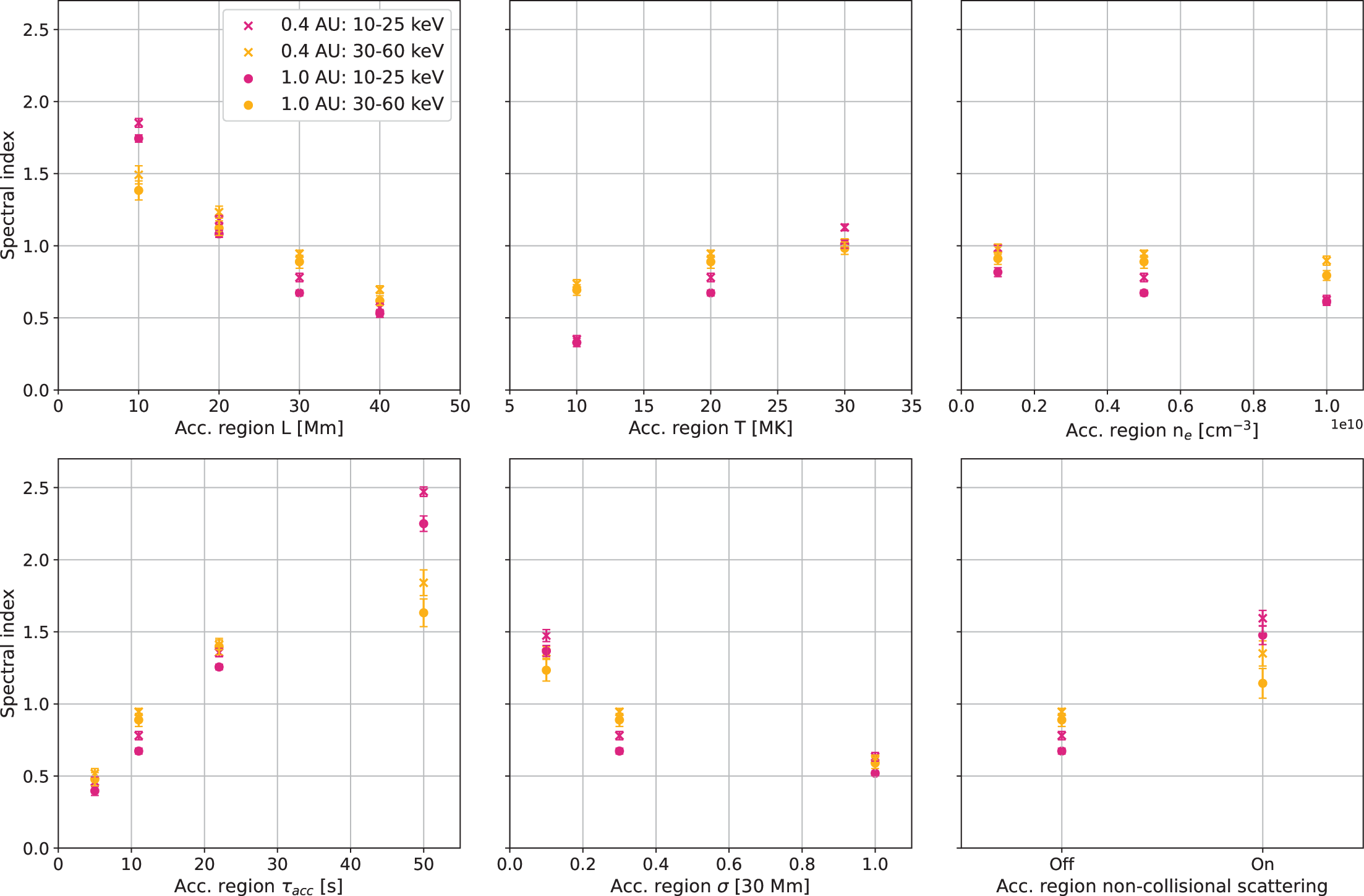}
    \caption{Spectral indices of the heliospheric electron peak flux distributions at 0.4 AU (crosses) and 1.0 AU (circles) in the absence of planetary magnetospheres. The electrons originate in a coronal region subject only to collisional and turbulent acceleration effects (other non-collisional effects are switched off by default), then ejected into a heliosphere subject only to adiabatic focusing and non-collisional scattering. Thermal and acceleration-profile parameters of the acceleration region are varied individually with other parameters fixed at intermediate values. The spectral index is shown for ranges of [10:25] keV (magenta) and [30:60] keV (yellow).}
    \label{fig:hel_spec}
\end{figure*}

Figure \ref{fig:hel_spec} shows the spectral indices of the peak electron fluxes at two points in the heliosphere, 0.4 AU and 1.0 AU, for varying values of acceleration region parameters. Increasing the size of the coronal acceleration region leads to a hardening of the spectra out to 1.0 AU for both the low and high energy ranges, with close alignment of the $10-25$ and $30-60$ keV indices across the heliosphere for a sufficiently large region. There is a significant difference between these indices for shorter regions and in the lower energy range, demonstrating a hardening of the spectra between 0.4 and 1.0 AU. Longer acceleration regions result in a greater degree of acceleration overall (as shown in Figure \ref{fig:pla_bound_params_comp}). The spectral index of the $10-25$ keV range shows a greater degree of hardening for longer acceleration regions than for higher energies. Also of note is the divergence between the high and low energy spectra indicating the presence of a spectral break where the region is short and acceleration is minimised.

A low temperature acceleration region shows hard spectra at the lower energy ranges as the thermal distribution is shifted to sub-5 keV energies and a significant hardening is present between $10-25$ keV compared to energies above this range. The spectral indices begin to align as the region temperature increases, due to a greater overlap of the thermal component with the measured spectral index ranges. The highest simulated temperature leads to no significant variation in the spectral index across the heliosphere for either energy range. As with $L$, the lower energy range is more sensitive to $T$ as that is where the thermal component begins to overlap when above 10 MK.

Higher-energy spectra indices remain relatively consistent across the heliosphere for all tested coronal number densities. For the lower-energy range there is a slight spectral hardening as density increases, though only to a minor degree. This suggests that heliospheric spectral indices are not particularly sensitive to acceleration region number densities, potentially as the increased collision rate disproportionately affects the sub-10 keV population and have a comparatively low impact on electrons at higher energies. As such, it may be difficult to constrain this parameter solely with in-situ spectra.

The acceleration profile parameters $\tau_{acc}$ and $\sigma$ demonstrate the same pattern in the in-situ spectra  as they do at the ejection boundary (see Figure \ref{fig:acc_bound_params_comp}), indicating that these features are well preserved through the non-collisional heliosphere. High-energy spectra appear harder than the low-energy range for longer $\tau_{acc}$, corresponding to less acceleration, though the same does not appear to be true for shorter values of $\sigma$. This would suggest that heliospheric spectra are more sensitive to the magnitude of turbulent acceleration than the physical extent of the profile, within the tested ranges.

Finally we examine how the spectral indices change when non-collisional effects (adiabatic focusing and scattering) are present in the acceleration region. In this case we take the mean free path $\lambda_s$ to be inversely proportional to the electron energy, as in Equation \ref{eq:lambda_acc}. We see a significant spectral softening at all energies, particular the lower range, consistent with Figure \ref{fig:NC_fluence_comp}b. As with the acceleration profile parameters, this indicates that these spectral features are well preserved in our heliospheric transport model.

\section{Paired heliospheric and chromospheric simulation}\label{sec:helio_plus_hxr}

To examine the observed properties of different populations of flare-accelerated electrons in our model, we now employ the model of e.g., \citet{2014ApJ...787...86J,2019ApJ...880..136J,2023ApJ...946...53S} to accelerate and transport an electron population ``at the Sun", i.e., an electron population accelerated within the closed magnetic topology of newly reconnected flaring coronal loops and then transported to denser layers of the solar chromosphere where the bulk of the electron energy is lost and bremsstrahlung X-rays are produced. For more details regarding this model, please see \citet{2023ApJ...946...53S}. This model uses an equivalent method as that outlined in this paper for accelerating electrons within a chosen coronal loop top region.

Initially, we accelerate electrons in both models using identical coronal plasma properties of $L = 20$~Mm, $T = 20$ MK, $n = 5 \times 10^9$ cm$^{-3}$ and acceleration properties of $\tau_{acc} = 11$s, $\chi = 3$, $\sigma = 0.3L$. For electrons at the Sun, we create the corresponding bremsstrahlung X-ray emission and plot the full-flare (coronal loop-top and footpoint sources) X-ray photon flux spectra observed by a satellite at Earth [photons s$^{-1}$ cm$^{-2}$ keV$^{-1}$], the prime diagnostic of flare-accelerated electrons at the Sun, as observed by previous and current instruments such as Reuven Ramaty High Energy Solar Spectroscopic Imager (RHESSI)  \citep{2002SoPh..210....3L} and SolO Spectrometer Telescope for Imaging X-rays (STIX) \citep{2020A&A...642A..15K}. 

\begin{figure*}[t!]
    \centering
\includegraphics[width=0.5\textwidth]{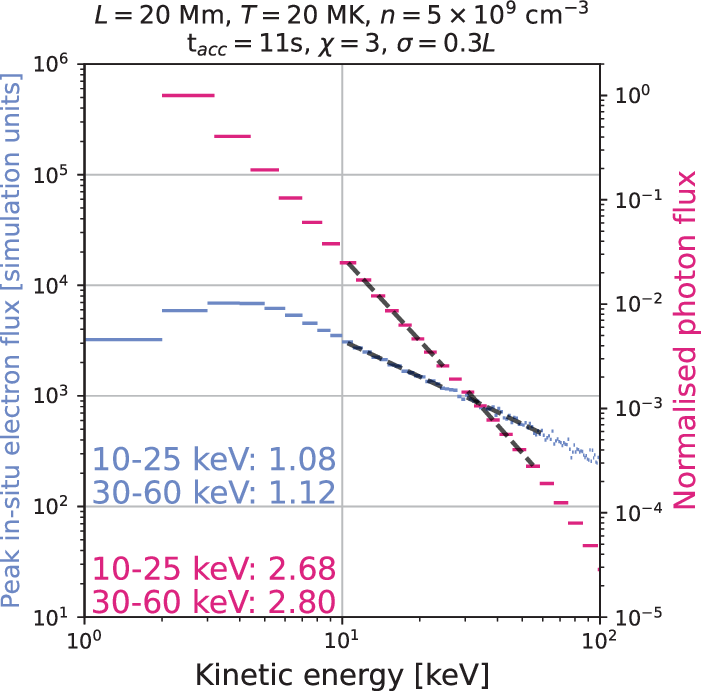}
    \caption{Comparison between simulated in-situ (1.0 AU) peak electron flux (blue) and simulated full-flare photon flux (red). The spectral indices between 10-25 and 30-60 keV have been calculated for both curves, with the power law fits over these ranges shown with dotted lines (black).}
    \label{fig:xray_comp}
\end{figure*}

The resulting outputs for both electron populations (X-ray emitting and heliospheric) are shown in Figure \ref{fig:xray_comp}. Here the simulated peak in-situ (1.0 AU) electron and full-flare photon flux curves are compared and their spectral indices extracted for the 10-25 and 30-60 keV energy ranges. For this case we find that the spectral indices of the in-situ peak electron flux ranges are $\delta = 1.08$ (10-25 keV) and $\delta = 1.12$ (30-60 keV). In contrast, the spectral indices of the full-flare photon flux are $\gamma = 2.68$ (10-25 keV) and $\gamma = 2.80$ (30-60 keV), which are softer than $\delta$ found in the heliosphere. This better matches the thin-target assumption used in \cite{2007ApJ...663L.109K} but does not match the approximate $\delta \approx \gamma$ relation found in the above study, with in-situ spectra found to be much harder than for the selected events. We note that in \cite{2023A&A...675A..27J} the sub-100 keV spectral indices for in-situ electrons measured by SolO Energetic Particle Detector (EPD) \citep{2020A&A...642A...7R} range between $\delta = 0.39 \pm 0.22$ and $\delta = 3.64 \pm 0.18$, indicating that harder spectra are observed in the heliosphere.

In order for the in-situ spectra to better match the correlation seen in \cite{2007ApJ...663L.109K}, there are several parameters that can be changed. The inclusion of non-collisional scattering demonstrates a softening of the spectra at 1.0 AU as does reducing $L$, $\tau_{acc}$ and $\sigma$, effectively reducing the degree of turbulent acceleration across the region. This would suggest that if electrons in both populations originate within a common source region, the acceleration and plasma properties may not be uniform across the extended region.

\section{Summary}

A parameter search was performed for the plasma and turbulent acceleration properties of a coronal acceleration region on the in-situ electron spectra, investigating how sensitive the spectral indices are to changes in these properties out to distances of 1.0 AU. We then compared a case of a common source acceleration region producing both in-situ and X-ray flux spectra.

\begin{itemize}
    \item Shorter acceleration timescales $\tau_{acc}$, greater velocity dependence $\chi$ and more extended acceleration profiles (i.e., larger $\sigma$) result in harder electron spectra at the acceleration boundary.
    \item Differences between the resultant spectral indices when these parameters are varied are preserved out to 1.0 AU.
    \item The in-situ spectral indices are also sensitive to the length of the acceleration region $L$, but are relatively unchanged subject to variation in region temperature $T$ and electron number density $n$.
    \item Electron spectra soften slightly as a result of adiabatic focusing in the corona.
    \item Short-timescale turbulent scattering results in a softening of the in-situ spectra by approximately 0.5.
    \item Comparing the in-situ electron and full-flare X-ray photon spectra produced by a common source region with shared plasma and turbulent acceleration properties shows a harder spectra for the in-situ electrons with $\delta \approx \gamma - 1.6$. Previous studies \citep[such as][]{2007ApJ...663L.109K} show an approximate $\delta \approx \gamma$ relation. Higher $\tau_{acc}$, shorter $\sigma$ and shorter $L$ in the in-situ model may replicate this relation, based on the trends discussed above.
\end{itemize}

In our next study we will use a similar method to compare with SolO/EPD and SolO/STIX flares. We also aim to improve our in-situ model with the implementation of a more realistic coronal-heliospheric environment model and the inclusion of wave-particle interactions, which have been shown to affect the spectra of low energy in-situ electrons \citep[e.g.][]{2009ApJ...695L.140K}.

\section*{Acknowledgments}
RP \& NLSJ gratefully acknowledge financial support from the Science and Technology Facilities Council (STFC) Grants ST/V000764/1 and ST/X001008/1. The authors acknowledge IDL support provided by STFC. NLSJ is supported by an international team grant \href{https://teams.issibern.ch/solarflarexray/team/}{“Measuring Solar Flare HXR Directivity using Stereoscopic Observations with SolO/STIX and X-ray Instrumentation at Earth}” from the International Space Sciences Institute (ISSI) Bern, Switzerland. MS is supported by the NASA Living with a Star Jack Eddy Postdoctoral Fellowship Program, administered by UCAR's Cooperative Programs for the
Advancement of Earth System Science (CPAESS) under award $\#$80NSSC22M0097. The data that support the findings of this study are available from the corresponding author upon reasonable request.

\bibliography{main}
\bibliographystyle{aasjournal}

\end{document}